\documentclass[twocolumn]{aa}
\usepackage{graphicx}
\usepackage{txfonts}
\begin{document}
   \title{I~Zw~1 observed with {\em XMM-Newton} }

   \subtitle{Low-energy spectral complexity, iron lines, and hard X-ray flares}

    \author{L. C. Gallo
        \inst{1}
        \and
        Th. Boller
        \inst{1}
        \and
        W. N. Brandt 
        \inst{2}
        \and
        A. C. Fabian 
        \inst{3}
	\and
        S. Vaughan   
        \inst{4}
          }

   \offprints{LCG (lgallo@mpe.mpg.de)}

   \institute{Max-Planck-Institut f\"ur extraterrestrische Physik, Postfach 1312, 85741 Garching, Germany
	 \and
    Department of Astronomy and Astrophysics, The Pennsylvania State University, 525 Davey Lab, University Park, PA 16802, USA
         \and
    Institute of  Astronomy, Madingley Road, Cambridge CB3 0HA, UK
         \and
    X-ray and Observational Astronomy Group, Department of Physics and Astronomy, University of Leicester, Leicester LE1 7RH, UK
             }

   \date{Received -- ; accepted -- }

   \abstract{
We present a 20 ks {\em XMM-Newton} observation of the prototypical Narrow-Line
Seyfert 1 galaxy I~Zw~1.  The best-fit model to the data
is a double blackbody plus a dominant power-law, on which complex soft absorption (possibly a
blended edge or absorption lines) and/or OVII emission are superimposed, as well as strong Fe K$\alpha$ emission.  The iron feature in the high-energy
spectra appears broad; however, on close examination of the EPIC pn data, there exists the possibility
that the broad emission feature can be attributed to a neutral Fe K$\alpha$ line in addition to a blend
of He- and H-like Fe K$\alpha$ lines.  
The light curve shows a strong, hard X-ray flare concentrated in the 3--12 keV band.
The flare appears to induce spectral variability, showing spectral hardening to be occuring as the flare intensifies.
A detailed examination suggests that the spectral variability is most likely due to an increase in the 3--12 keV 
flux relative to the soft flux during the flare.
A difference spectrum and complete modelling of the flare and non-flare spectra show intrinsic changes 
only in the normalisation of the continuum components and not in their shape parameters.
The timing results are consistent with the flare originating in the accretion disc corona.
The iron emission line(s) do not appear to respond to changes in the continuum flux during the flare; the iron lines
are stronger in equivalent width during the low-flux (non-flare) states, and weaker during the flare. 

   \keywords{galaxies: active --
        galaxies: individual: I~Zw~1 --
        X-rays: galaxies
               }
   }

   \maketitle
%

\section{Introduction}
Aside from being the prototype Narrow-Line Seyfert 1 galaxy (NLS1), I~Zw~1 
is an important object having influence in many areas of AGN astronomy.  Its strong and
narrow optical/UV emission lines minimize blending effects and allow I~Zw~1 to be used as a template for
optical (Boroson \& Green 1992) and UV (Vestergaard \& Wilkes 2001; Laor et al. 1997)
FeII emission.  Taking its luminosity into consideration (M$_B$ $\sim$ $-$23.5; Schmidt \& Green
1983), I~Zw~1 is also defined as a quasar, likely making it the nearest quasar known
($z$ = 0.0611).  Although its infrared luminosity falls just short of those of 
Ultraluminous Infrared galaxies (ULIRGs), I~Zw~1 exhibits properties of both QSOs and ULIRGs, and hence
Canalizo \& Stockton (2001) consider it as a possible transition object from QSO to ULIRG.
Scharw\"achter et al. (2003; see also Eckart et al. 2000) used VLT observations to examine the host 
properties of I~Zw~1.  They observe an interacting spiral galaxy with a young stellar population 
in the 
disc as well as young, hot stars, and supergiants in the bulge, which is indicative of strong starburst activity.  

At X-ray energies, I~Zw~1 displays many of the characteristics typically associated with 
NLS1.  $ROSAT$ observations (Boller et al. 1996; Lawrence et al. 1997) found a steep 0.1--2.4 keV 
spectrum, described by 
a power-law of photon index $\sim$3, and notable variability (change by a factor of 1.5 in $\sim$6000 s).
However, Leighly (1999a) found that an absorbed power-law, without a soft excess, was sufficient to 
fit the 0.5--10 keV $ASCA$ continuum. 
In addition, Leighly detected a very strong ($EW$ $\approx$ 1300 eV) Fe K$\alpha$ line (see also Reeves \& Turner 2000), 
which suggested a nucleus-wide 
overabundance of iron in I~Zw~1.

In this paper we present a 20 ks {\em XMM-Newton} observation of I~Zw~1.
We begin by describing the observation and the data reduction in the next section.  Section 3
is devoted to the results from the Optical Monitor.  In Section 4 we discuss the X-ray spectra observed with the EPIC. 
In Section 5 we concentrate on the timing properties.  Lastly, in Section 6, we will
conclude with a summary of our results.

\section{Observation and data reduction}
I~Zw~1 was observed with {\em XMM-Newton} (Jansen et
al. 2001) on 2002 June 22 during revolution 0464 for about 20 ks.
During this time all instruments were functioning normally.  The EPIC
pn camera (Str\"uder et al. 2001) was operated in large-window mode, and
the two MOS cameras (MOS1 and MOS2; Turner et al.  2001) were operated
in small-window mode.  All of the EPIC cameras used the medium filter.
The two Reflection Grating Spectrometers (RGS1 and RGS2; den Herder et
al. 2001)
and the Optical Monitor (OM; Mason et al. 2001)
also gathered data during this time. The Observation Data Files (ODFs)
were processed to produce calibrated event lists using the {\em
XMM-Newton} Science Analysis System ({\tt SAS v5.4.1}). Unwanted hot,
dead, or flickering pixels were removed as were events due to
electronic noise.  Event energies were corrected for charge-transfer
losses.  EPIC response matrices were generated using the {\tt SAS} tasks
{\tt ARFGEN} and {\tt RMFGEN}. 
Light curves were extracted from these event lists to
search for periods of high background flaring.  A short (lasting $<$ 300 s),
small-amplitude flare (equivalent to less than 3\% of the source count rate) was
detected about 9.5 ks into the observation.  The data for this time period
were ignored during the analysis.
The total amount of good exposure
time selected was 18650 s and 21006 s for the pn and MOS detectors,
respectively. The source plus background photons were extracted from a
circular region with a radius of 35 arcsec, and the background was
selected from an off-source region and  appropriately scaled to the source
region.  Single and double events were selected for the pn
detector, and single-quadruple events were selected for the MOS. 
The statistics were clearly dominated by the source at energies below 10 keV.
The total numbers of good counts collected by the pn, MOS1, and MOS2 instruments
were 137840, 45581, and 46829, respectively.  The {\em XMM-Newton} observation provides 
a substantial improvement in spectral quality over the 28.3 ks exposure (92.4 ks duration)
$ASCA$ observation in which $\approx$12000 counts were collected (Leighly 1999b).

High-resolution
spectra were obtained with the RGS.  The RGS were operated
in standard Spectro+Q mode for a total exposure time of 20713 s.  The
first-order RGS spectra were extracted using {\tt RGSPROC}, and the response
matrices were generated using {\tt RGSRMFGEN}.

The Optical Monitor collected data through the UVW2 filter
(1800--2250 \AA) for about the first 9 ks of the observation, and then it was switched
to grism mode for the remaining time.

\section{UV analysis}

The Optical Monitor collected data in the fast mode through 
the UVW2 filter (1800--2250 \AA) for about the first half of the observation.
The rest-frame wavelength range observed through the UVW2 filter is roughly 1700--2120 \AA.
Besides continuum emission, Laor et al. (1997) attributed much of the emission in this 
spectral region to FeII and FeIII multiplets.  There is also some strong emission 
from AlIII, and semi-forbidden species of SiIII], CIII], and NIII]. 

In total, five photometric images were taken, each exposure lasting 1660 s.
The apparent UVW2 magnitude is 14.36 $\pm$ 0.04 corresponding to a flux density of
9.94 $\times$ 10$^{-15}$ erg s$^{-1}$ cm$^{-2}$ \AA$^{-1}$ 
which is about a factor
of three lower than during the 1994 $HST$ observation (Laor et al. 1997).
The approximately 2.5 hour light curve is flat with the exception of one data point
which exhibits a 1$\sigma$ deviation from the average.  We do not consider this 
event significant.

Taking advantage of the simultaneous UV/X-ray observations, we derive the nominal spectral 
slope, $\alpha_{ox}$, between 2500 \AA~and 2 keV.  For the UV flux, we use the peak response
of the UVW2 filter (2000 \AA~in the rest-frame).  We note 
from the $HST$ spectrum, that the flux density at 2000 \AA~is approximately equal to the flux density
at 2500 \AA~(Laor et al. 1997).  By assuming a negligible continuum slope between 2000--2500 \AA~we
are able to extrapolate the UVW2 flux at 2000\AA~to 2500 \AA.
For the model
dependent X-ray flux (see Section 4) we used only the EPIC data which were collected
simultaneously with the UV data (i.e. the first 10 ks of the observation).
We determine a value of $\alpha_{ox}$ = 1.20 which is notably X-ray stronger than the
value of 1.41 reported by Lawrence et al. (1997), and 1.56 reported by Brandt et al. (2000).\footnote{
Brandt et al. (2000) use a reference point of 3000 \AA~for the UV flux.  It is also worth noting that at 3000 \AA~there 
is a relatively strong FeIII multiplet emission that will make I~Zw~1 appear X-ray $weaker$ than if we
were to consider only the continuum.}  
The spectral index between 2500 \AA~and 2 keV is clearly 
variable with time, but always appears X-ray strong compared to the predicted value for radio-quiet quasars of this luminosity
(Vignali et al. 2003).

\section{X-ray spectral analysis}

The source spectra were grouped such that each bin contained at least 40 counts. Spectral fitting
was performed using {\tt XSPEC v11.2.0} (Arnaud 1996).  Fit parameters are reported in the
rest-frame of the object, although the figures remain in the observed-frame.  
The quoted errors on the model parameters correspond to a 90\% confidence level for one interesting
parameter (i.e. a $\Delta\chi^2$ = 2.7 criterion),
and luminosities are derived assuming isotropic emission. 
The Galactic column density toward I~Zw~1 is $N_H$ = 5.07 $\times$
10$^{20}$ cm$^{-2}$ (Elvis et al. 1989).
A value of the Hubble constant of $H_0$=$\rm 70\ km\ s^{-1}\ Mpc^{-1}$
and a WMAP-cosmology with $\Omega_{M}$ = 0.3 
and $\Omega_\Lambda$ = 0.7 have been adopted throughout.

Initially all three EPIC spectra were fitted separately to examine any discrepancies.
The 3--10 keV spectra were first fitted with a single power-law modified by
Galactic absorption. 
The pn and MOS2 are entirely consistent; the photon indices for the pn and 
MOS2 fits are 2.26$^{+0.05}_{-0.06}$ and 2.18 $\pm$ 0.10, respectively.
However, the MOS1 photon index is much flatter ($\Gamma_{MOS1}$ = 1.95 $\pm$ 0.10)
and does not lie within the 90\% confidence range of the other EPIC instruments.
This inconsistency in the MOS1 data was previously realised in
an observation of 3C~273 by Molendi \& Sembay (2003).
Molendi \& Sembay noted a difference in the MOS1 photon index compared to the other EPIC photon 
indices of $\Delta\Gamma$ $\sim$ 0.1.  The difference in our observation is larger 
(0.13 $<$ $\Delta\Gamma$ $<$ 0.31).  
However, the pn data dominate the fits above 3 keV given the higher statistics afforded by this instrument. 
Including the 3--10 keV
MOS1 data has little influence on the model results.
For clarity, the figures are shown without the 3--10 keV MOS1 data, but the fit statistics
will include all of the data.
Cross-calibration uncertainties between pn and MOS appeared quite significant in our observation
at energies below $\sim$ 0.7 keV; therefore, the EPIC data are only fitted together at energies above
0.7 keV.  Between 0.7--3 keV, the data from all three instruments appears entirely consistent.

\begin{figure}
\includegraphics[width=6.0cm,angle=-90,clip=]{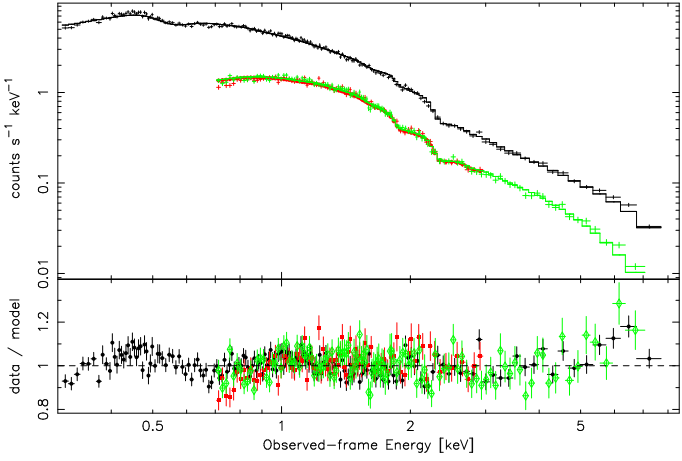}
\caption{In the upper panel we present an absorbed power-law fitted to the EPIC 0.3--10 keV data. The
top curve is the pn data and the data points in the bottom curve are the MOS1 and MOS2 data.  In the lower
panel are the residuals (data/model) of the power-law fit.  Black circles, red squares, and green diamonds are the
pn, MOS1, and MOS2 residuals, respectively. The data have been highly binned for display purposes only.
}
\label{pofit}
\end{figure}

In Figure~\ref{pofit} we present the 0.3--10 keV EPIC spectra fitted with an 
absorbed power-law.  The fit is reasonable given its simplicity ($\chi^2$ = 1325.4/1076 $dof$); 
however, a number of features are immediately apparent. 
The total photoelectric absorption is nearly twice that expected from 
the Galactic hydrogen column density, indicating that I~Zw~1 is intrinsically absorbed.
An excess above the power-law is observed between 0.4--0.6 keV.  The weak soft excess is marked on the 
red side by absorption which extends to the energy limit of the detector, and on the blue side by an
absorption feature which dips below the power-law from 0.6--0.8 keV.
Positive residuals between 6--7 keV likely suggest emission due to iron.

\subsection{Fe K emission}
We begin the spectral analysis by considering the 3--10 keV band alone.
Fitting a power-law in this region results in positive residuals between
6--7 keV in the observed-frame, as is shown in Figure~\ref{pofit} ($\chi^2$ = 333.4/295 $dof$).  In addition, the fit demonstrates that the spectrum
is steep compared to broad-line Seyfert 1s, with $\Gamma$ = 2.24 $\pm$ 0.05.
We attempt several model fits to describe the feature between
6--7 keV.  A Gaussian model, diskline model (Fabian et al. 1989), and a Laor line model (Laor 1991)
are all relatively successful in fitting the data, but none could be ruled superior over
the others.  In all cases, the addition of a line model results in further steepening of the 
power-law continuum ($\Gamma$ $\sim$ 2.3).  
The best-fitting parameters to the simplest line model (i.e. Gaussian profile) are $E$ = 6.69$^{+0.14}_{-0.33}$ keV,
$\sigma$ = 316$^{+268}_{-136}$ eV, and $EW$ = 179$\pm$17 eV ($\chi^2$ = 309.5/292 $dof$).

\begin{figure}
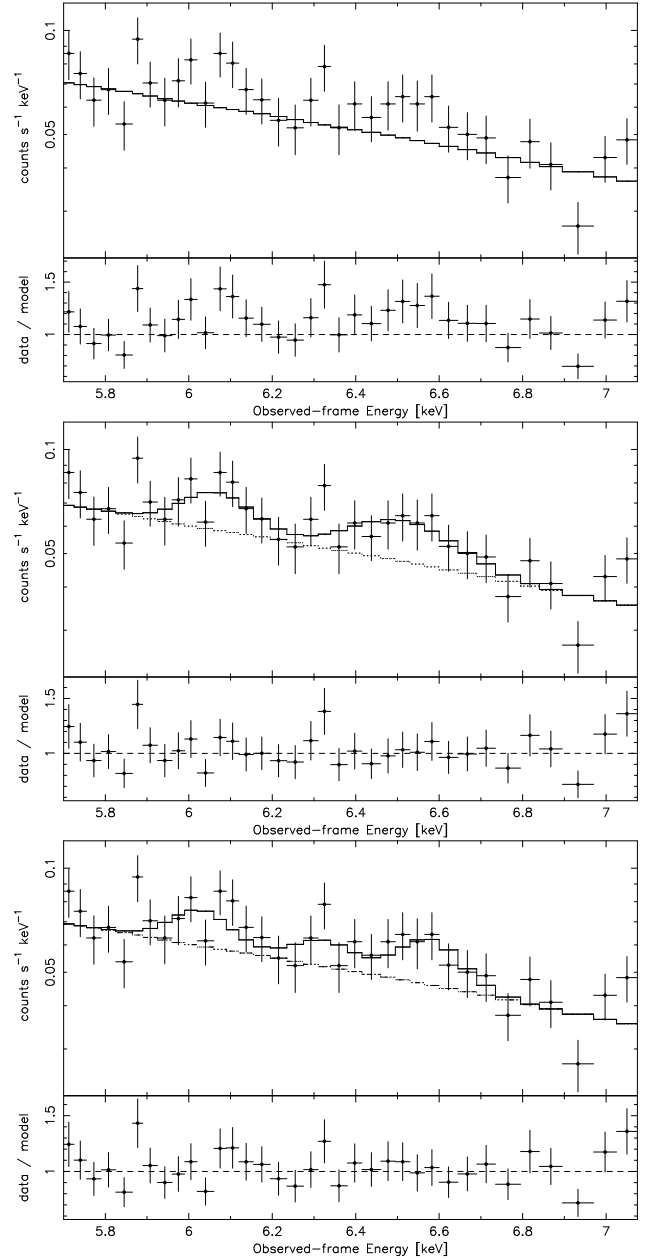

\includegraphics[angle=-90,width=8.3cm,clip=]{0411f2a.ps}  
\includegraphics[angle=-90,width=8.3cm,clip=]{0411f2b.ps}
\includegraphics[angle=-90,width=8.3cm,clip=]{0411f2c.ps}
\caption{Spectral fit to the pn data between 5.7--7.1 keV.
{\bf Upper panel}:  The data are fitted with the best-fit 3--10 keV power-law.  Two separate ``humps''
are clearly seen in the residuals.
{\bf Middle panel}:  Two Gaussian profiles are applied to the fit.  The first line represents emission
from neutral Fe K$\alpha$ ($E$ = 6.4 keV), and the second line
($E$ $\approx$ 6.9 keV) is consistent with emission from a blend of 
H- and He-like iron.
{\bf Lower panel}:  Three intrinsically narrow Gaussians are fitted to model possible emission from
neutral ($E$ = 6.4 keV), He-like ($E$ = 6.7 keV), and H-like ($E$ = 6.97 keV) Fe K$\alpha$.
}
\label{lines}
\end{figure}

A further intriguing possibility for the reflection component is the existence of multiple emission lines.
In Figure~\ref{lines}, we have displayed the 5.7--7.1 keV pn data (6.0--7.5 keV rest-frame energies), to take advantage of
the higher signal-to-noise data afforded by the EPIC pn instrument.
On careful inspection of the residuals in the pn data we can distinguish two separate 
humps between 6--7 keV.  We introduce a single Gaussian into the fit and fix its energy at 
6.40 keV to account for emission from neutral Fe K$\alpha$.
The resulting line width is $\approx$ 70 eV, and both the quality of the fit and the residuals are improved 
($\Delta\chi^2$ = 20.2 for 2 additional free 
parameter).
A second Gaussian is added to model the remaining residuals at slightly higher energies.  This second Gaussian
further improves the fit and minimises the residuals in this spectral region
($\Delta\chi^2$ = 6.4 for 3 additional free parameters).  This second line has an energy of 
$E$ $\approx$ 6.87 keV and it is slightly broader 
($\sigma$ $\approx$ 140 eV) than the neutral Fe K$\alpha$ line. 
The energy of this broad line
is consistent with emission from a blend of H- and He-like iron.
Formal modelling, with all parameters left free to vary, 
resulted in a similar fit quality as the {\em step-wise} process just described. 
($\chi^2$ = 307.4/290 $dof$).  

We further examined the possibility that the positive residuals between 6--7 keV could be the result of three narrow Fe K$\alpha$ lines:
neutral (6.40 keV), He-like (6.70 keV), H-like (6.97 keV) lines.  We included three narrow ($\sigma$ = 10 eV) Gaussian
profiles at the above stated energies and computed the fit.  The triple line model is a good fit to the EPIC data
($\chi^2$ = 308.6/293 $dof$).  We obtain values of $EW$ = 32 $\pm$ 1eV, 52 $\pm$ 2 eV, and 52 $\pm$ 2 eV for the neutral,
He-like, and H-like Fe K$\alpha$ lines, respectively.  

Modelling of the $ASCA$ observation of I~Zw~1 required a strong, (partially) ionised Fe K$\alpha$ line.  Leighly (1999a) fitted 
a 1300 eV line with a rest-energy of 7.0$^{+0.6}_{-0.3}$ keV, while Reeves \& Turner (2000) established the line centre at 6.77$^{+0.11}_{-0.17}$ keV
with $EW$ = 483$^{+212}_{-211}$ eV.  Both line energies are consistent with our simultaneous fit of the pn and MOS data, using a single, broad 
Gaussian profile ($E$ = 6.69$^{+0.14}_{-0.33}$ keV), though our line appears substantially weaker ($EW$ = 179 $\pm$ 17 eV). 
Utilizing the higher sensitivity of the EPIC pn detector, we find that the broad feature between 6--7 keV can be well described by emission from
multiple narrower lines, 
showing that the evidence for a truly broad line is weak.  
Statistically there is little difference between the two and three Gaussian fits, both are acceptable; however, the residuals at about 6.7 keV
($\sim$6.3 keV in the observed frame) are slightly improved by the three narrow Gaussians model.   
A longer exposure of I~Zw~1 with {\em XMM-Newton}, in order to boost the signal-to-noise of the MOS data, would clarify the situation
in this spectral region.

\subsection{The broad-band spectrum}

As is illustrated by Figure~\ref{pofit} a power-law model is a reasonable starting point for the broad-band (0.3--10 keV)
fit.  With the exception of some complexities at energies lower than $\sim$ 1 keV, the spectrum is relatively smooth.
For simplicity, we shall fit the iron complex (6--7 keV) with a double Gaussian model: the first Gaussian profile will
be fixed at 6.4 keV to describe emission from neutral iron, and the second profile at slightly higher energies will be 
used to approximate emission from ionised species of iron.  
In addition, we have introduced a second cold absorption term.  The first term is fixed to the Galactic value, while
the second term will be allowed to vary and will be used to estimate the neutral absorption in I~Zw~1.

With these additional parameters in place, the single power-law fit was improved at high energies, but the
residuals below $\sim$ 1 keV were not affected ($\chi^2$ = 1330.7/1071 $dof$).  We added a second continuum component to the 
existing model to fit the weak soft excess at lower energies.  Neither a single black body ($\Delta\chi^2$ = 7 for the addition of 2 free parameters),
nor a second power-law ($\Delta\chi^2$ = 10 for the addition of 2 free parameters)
were entirely effective in fitting the soft excess.
A significant improvement in the overall fit is seen when the single power-law continuum is replaced with a broken
power-law ($\Delta\chi^2$ = 67.4 for the addition of 2 free parameters).  In this model the hard power-law ($\Gamma$ $\approx$ 2.48) dominates
the spectrum at energies higher than $\approx$0.75 keV.  At lower energies the spectrum is fitted with a steeper power-law
with a photon index of ($\Gamma$ $\approx $2.86.  An $F$-test suggests that the broken power-law fit is an improvement over
the single power-law at $>99.99$\% confidence, and that a second softer component is a required addition.
An alternative model, which is thermal in nature, is a double blackbody fit ($\Delta\chi^2$ = 98.3 for the addition of 4 free parameters).
In this case, the steep power-law ($\Gamma$ $\approx$ 2.32) continues to dominate much of the spectrum; however,
the cooler black body ($kT_1$$\approx$78 eV) contributes about 20\% of the total unabsorbed flux, while the hotter black body
($kT_2$$\approx$213 eV) contributes about 5\%.

\begin{figure}
\includegraphics[width=6.4cm,angle=-90,clip=]{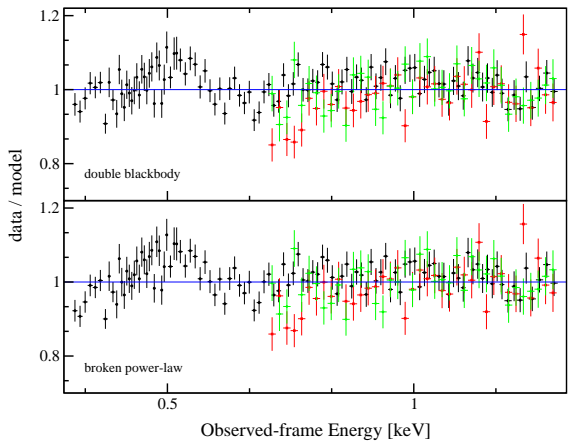}   
\caption{In the top panel we show the ratio (data/model) of the absorbed
double blackbody plus power-law fit to the 0.3--1.3 EPIC data.  The lower panel
is the absorbed broken power-law fit.  Although both are an improvement over a single power-law,
neither model is able to minimise the residuals  
(see text for details).  The black circles are the pn data; the red squares and
green diamonds are the MOS1 and MOS2 data, respectively.  The data have been highly binned for display purposes only.
}
\label{softrat}
\end{figure}

Even though both the broken power-law and double blackbody models are dramatic improvements over the single power-law fit
statistically, the residuals in the low-energy band are still unsatisfactory (Figure~\ref{softrat}).
In both cases, the residuals in the observed frame between 0.4--0.6 keV still exhibit the up-down trend seen in the single power-law
fit, while between 0.6--0.75 keV
the data remain slightly overestimated.  
Unfortunately, the short exposure and low signal-to-noise in the RGS does not allow us to examine productively the 
soft spectrum of I~Zw~1.  A {\em by-eye} inspection of the RGS spectrum does indicate a number of possible narrow absorption and emission features, but formal modelling
of the spectrum does not reveal any significant results.

One successful attempt to minimise the residuals in the soft spectrum is by the addition
of a Gaussian emission line.  Depending on the continuum model used, the line energy is between 535--565 eV, with a width of
$\sigma$ = 40--55 eV and an equivalent width of 25--30 eV.  The addition of the soft emission feature to the broken
power-law ($\Delta\chi^2$ = 84 for the addition of 3 free parameters) or double blackbody model ($\Delta\chi^2$ = 65.6 for the addition of 3 free parameters) is significant.
The measured energy is consistent with emission from the He$\alpha$ triplet of OVII at $\sim$22 \AA~(the resonance,
intercombination, and forbidden lines), which are known to be strong in some Seyfert galaxies (e.g. NGC~3783,
Kaspi et al. 2002; Behar et al. 2003).  At least two ``emission line-like'' features are suggested by the RGS spectrum
at this energy, giving credence to this model.  
As a word of caution, we would like to make note of the calibration issues around the O-edge in the EPIC pn detector.
Although the calibration is significantly improved in {\tt SAS v5.4.1} (Kirsch 2003), and we do not believe that it has
any bearing in this observation, it is difficult to make strong claims about this feature without independent data. 

An alternative method to minimise the residuals in the soft spectrum is to add an absorption feature (edge or line)
on the blue-side of the excess.
I~Zw~1 does possess UV absorption (Laor et al. 1997; Crenshaw et al. 1999), so it seems perfectly reasonable that it would also
possesses some complex X-ray absorption.  

In the first attempt we introduce an absorption line into the fit.
The addition of the absorption feature results in a small statistical improvement to the broken power-law
model ($\Delta\chi^2$ = 18.7 for the addition of 3 free parameters).  The line energy is at $E$ $\approx$ 749 eV and it is very weak ($EW$ $\approx$ $-$6 eV).
However, the addition of the absorption line is better received by the double blackbody model ($\Delta\chi^2$ = 49.9 for the addition of 3 free parameters).  
In this case, the equivalent width is about $-$22 eV and the line is centred at about 737 eV.  The addition of the line also
minimises the residuals in a satisfactory manner.  Broad absorption features at these energies are most consistent with
unresolved transition array (UTA) absorption due to Fe M-shell ions ($\sim$729--774 eV; Behar et al. 2001),
as observed in IRAS 13349+2438 (Sako et al. 2001), NGC 3783 (Blustin et al. 2002; Kaspi et al. 2002; Behar et al. 2003), and Mrk 509 (Pounds et al. 2001).

Alternatively, the low-energy absorption line can be replaced with an absorption edge.  Again, for the two continuum models
we are working with, the statistics are improved: $\Delta\chi^2$ = 68.4 and 60 for the addition of 2 free
parameters for the broken power-law fit
and the double blackbody fit, respectively.  For both models, the optical depth of the edge is $\tau$ $\approx$ 0.23--0.26, and the 
edge energy is at $E$ $\approx$ 664--689 eV.  The best-fit edge energies do not agree with the strongest edge feature which is expected
at 731 eV (OVII).  However, this inconsistency in the edge energy alone does not permit us to discard this interpretation.
The edge could very well be blended with other absorption features, perhaps even with some neutral iron absorption as was seen in
MCG--6--30--15 (Lee et al. 2001), in which case our model is too simple.

\begin{table*}
\begin{center}
\begin{tabular}{lcccccccccc}
(1) & (2) & (3) & (4)  & (5) & (6) & (7) & (8) & (9) & (10) & (11)  \\
Model & $\chi^2_\nu$ & $N_{IZw1}$ & Component 1 &  Component 2 & Component 3 & $E_{edge}$ & $\tau$ & $E_{line}$ & $\sigma$ & $EW$   \\
\hline
(a) & 1.10 (1065) & 10.1$\pm$0.2 & 198$\pm$5 & 94$\pm$2 & 2.35$\pm$0.01 & 664$^{+10}_{-13}$ & 0.26$^{+0.02}_{-0.01}$ & 6.4$^f$ & 43$^b$ & 38$^b$ \\
    &             &              &           &          &               &      &      & 6.84$^{+0.09}_{-0.11}$ & 188$^{+140}_{-90}$ & 128$^b$ \\
(b) & 1.11 (1067) & 10.0$\pm$0.1 & 2.80$\pm$0.02 & 1.44$^{+0.06}_{-0.05}$ & 2.47$\pm$0.02 & 689$^{+9}_{-13}$ & 0.23$^{+0.02}_{-0.01}$ & 6.4$^f$ & 1572$^{+616}_{-400}$ & 593$^b$ \\
    &             &              &           &          &               &      &      & 6.73$^{+0.17}_{-0.18}$ & 227$^{+177}_{-154}$ & 92$^b$ \\ 
(c) & 1.10 (1064) & 10.8$\pm$0.1 & 177$\pm$5 & 67$\pm$2 & 2.35$\pm$0.01 & -- & -- & 6.4$^f$ & 47$^b$ & 41$^b$ \\
    &             &              & &     &    &      &  & 6.84$^{+0.08}_{-0.11}$ & 187$^{+114}_{-69}$ & 135$^b$ \\
    &             &              & &     &    &      &  & 0.56$\pm$0.01 & 45$^b$ & 28$^b$ \\
\hline

\end{tabular}
\caption{Best-fit model parameters to the 0.3--10 keV EPIC data. 
 The columns are:
(1) the applied model (in addition to two FeK$\alpha$ emission lines): (a) double blackbody continuum with absorption edge; (b) broken power-law continuum with absorption edge; (c) double blackbody continuum with soft X-ray emission line; (2) reduced $\chi^2$ with $dof$ in brackets, 
(3) intrinsic photoelectric absorption ($\times$10$^{20}$cm$^{-2}$), (4) (a \& c) $kT_1$ (eV); (b) $\Gamma_1$, (5) (a \& c) $kT_2$ (eV); (b) break energy (keV), 
(6) (a \& c) $\Gamma$; (b) $\Gamma_2$, (7) edge energy (eV); (8) edge optical depth, (9) line energy (keV), (10) line width (eV), and (11) line equivalent width (eV).
The superscript $f$ indicates that the parameter is fixed to that value, and
$b$ indicates only the best-fit value.  
}
\end{center}
\end{table*}

The three best-fitting phenomenological models of the 0.3--10 keV spectrum of I~Zw~1 are shown in Table 1.  The continuum can be
modelled by either a broken power-law or a double blackbody and power-law.  There are primarily two difficulties in distinguishing between these 
apparently very different models.  The first is that the soft spectrum is complex.  We demonstrated that an absorption
edge or an emission feature is a significant improvement to both continuum models, in addition to extra neutral absorption.  The second fact is that the 
soft excess is weak, and the hard power-law component dominates much of the spectrum down to $\sim$750 eV.

All of the models have the disadvantage of possibly overestimating the photoelectric absorption in I~Zw~1.  The models require cold absorption on
the order of 10$^{21}$ cm$^{-2}$ in addition to the Galactic value, which is nearly a factor of 10 higher than was estimated
by the $ROSAT$ observations (Boller et al. 1996; Lawrence et al. 1997).  Given that $ROSAT$ was sensitive down to 0.1 keV,
it stands to reason that its value for the intrinsic absorption is reliable.
However, the high value measured with {\em XMM-Newton} is most likely a proxy for complicated absorption occuring below 0.4 keV
(e.g. blended absorption lines and/or edges).  
Given the very complex nature of the low-energy absorption in Seyferts, fitting simple models such as these can only be
illustrative of a much more complicated situation.  A long exposure, to utilise the high-resolution RGS, is required to 
probe the soft absorption in a meaningful way.

\begin{figure}
\includegraphics[width=6.4cm,angle=-90,clip=]{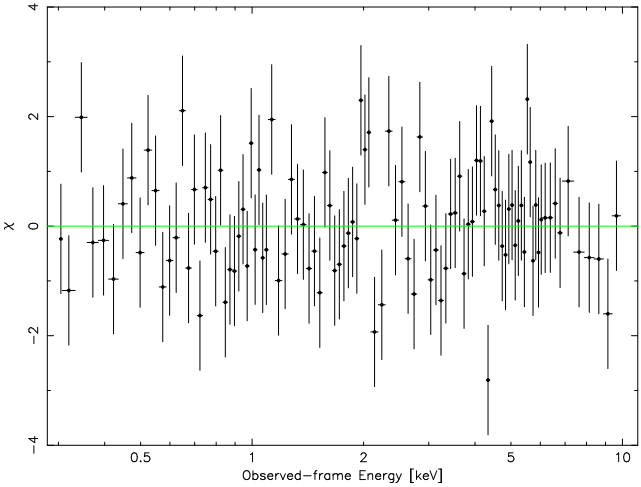}
\caption{The broad-band data-to-model residuals (in terms of $\sigma$) obtained from the double blackbody
phenomenological
fit to the EPIC data (see text and Table 1 for details).
For clarity, only the highly-binned pn residuals are shown.
}
\label{phenom}
\end{figure}

A slight advantage is given to the double blackbody plus power-law continuum in its ability to fit the iron emission 
between 6--7 keV.
The model is successful in predicting a high-energy continuum and an iron complex which is similar to that described in 
Section 4.1.  In contrast, the
broken power-law continuum requires a very broad ($\sigma$$\approx$1.5 keV) neutral line.  The neutral line width is certainly overestimated
by the fit to compensate for a continuum which is poorly defined between 7--10 keV.
A possible shortcoming for the double blackbody model is that the temperatures estimated for the thermal components
are high ($kT_1$$\approx$198 eV and $kT_2$$\approx$94 eV) for a standard accretion disc around a 10$^7$ M$_\odot$ black hole.  
However, since we do not understand the physics of the soft excess very well 
we cannot automatically deem the model as unphysical.  For example, bulk motions in the disc (Socrates et al. 2003)
is just one way
in which the shape of the soft excess can be altered from that expected from a standard accretion disc.
If we allow a soft excess component with curvature (rather than a power-law) the systematic residuals at low
energies can be made to diminish.  In addition, such curvature in the soft excess allows for better fitting of the high-energy continuum component and hence, the iron emission features. 
The final model residuals are shown in Figure~\ref{phenom}.

The average 0.3--10 keV flux, estimated from the double blackbody fit, and correcting for Galactic extinction is
$\sim$2.40 $\times$ 10$^{-11}$ erg s$^{-1}$ cm$^{-2}$.  The corresponding luminosity is 2.03 $\times$ 10$^{44}$ erg s$^{-1}$,
and the luminosity in the 2--10 keV band is 0.72 $\times$ 10$^{44}$ erg s$^{-1}$.  In comparison,
during the $ASCA$ observation the 0.5--10 keV luminosity was 1.36 $\times$ 10$^{44}$ erg s$^{-1}$, and the 
2--10 keV luminosity was 0.53 $\times$ 10$^{44}$ erg s$^{-1}$ (Leighly 1999).

\section{Timing analysis}
In this section, we will discuss the variability properties of I~Zw~1.
The pn and MOS light curves were both analyzed and found to be entirely consistent.
We will only discuss the results from the pn, given its higher signal-to-noise.
The 1$\sigma$ errors on the light curves were calcluated using counting statistics.

\subsection{The broad-band light curve}
\begin{figure}
\includegraphics[width=6.4cm,angle=-90,clip=]{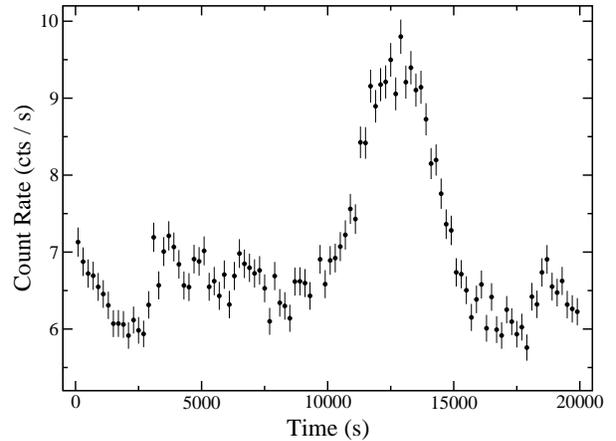}
\caption{The 0.3--12 keV light curve of I~Zw~1 in 200 s bins.
Zero seconds on the time axis marks the start of the observation at 09:18:44 on 2002--06--22.
}
\label{lc}
\end{figure}

In Figure~\ref{lc} we present the 0.3--12 keV EPIC pn light curve of I~Zw~1 in 200 s bins.  
The average count rate is 6.97 $\pm$ 0.19 counts s$^{-1}$.  The minimum and maximum 0.3--12 keV
count rates are 5.76 $\pm$ 0.17 and 9.80 $\pm$ 0.22 counts s$^{-1}$, respectively.
The most notable feature in the light curve is obviously the flare starting at 
$\sim$ 10 ks, when the count rate increases by $\sim$40\% in 2500 s.  The flare is remarkably
symmetric as the count rate returns from maximum to its pre-flare value in the same amount of time.
Except for the flare, the rest of the light curve is surprisingly serene. 
Considering only the periods between 0--10 ks and 15--20 ks, 
the fluctuations about the mean are only about 10\%, which is comparable to the observed variability on similar
time scales during
the $ASCA$ observations (Leighly 1999b); but modest compared to many NLS1 X-ray light curves. 
However, an observation with $ROSAT$ did find that the 0.1--2.4 keV flux dropped by about 25\% in $\sim$6000 s
(Boller et al. 1996).  It was not clear from that observation if the flux drop was subsequent to a rapid rise or 
part of a gradual decrease for an extended period of time.
In the remainder of this paper we will often
make reference to I~Zw~1 in its high- and low-flux states.  In defining the high-flux state, we
have selected the entire flaring event from $\sim$ 10--15 ks.  The low-flux state includes all of
the data from the remaining time intervals.  

We have calculated the radiative efficiency ($\eta$), assuming photon diffusion through a spherical 
mass of accreting material (Fabian 1979; but see also Brandt et al.
1999 for a discussion of important caveats).  The most rapid rise occurs from 10.1--12.6 ks
when the luminosity increases by 7.41 $\times$ 10$^{43}$ ergs s$^{-1}$.  The corresponding
radiative efficiency is only $\eta$ $\ge$ 1.5\%, which does not necessarily require  
radiative boosting or anisotropic emission during the flaring event.

\begin{figure}
\includegraphics[width=6.4cm,angle=-90,clip=]{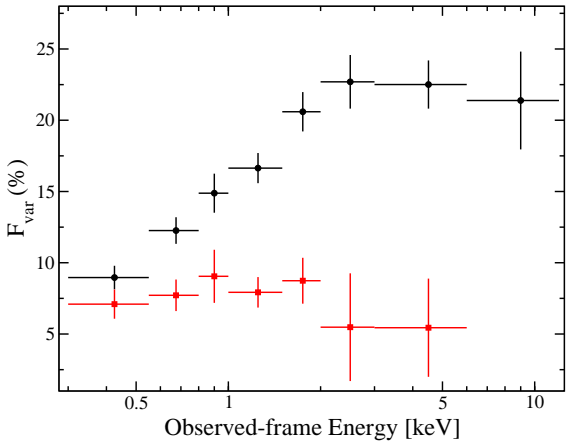}
\caption{F$_{var}$ calculated in eight energy bins between 0.3--12 keV using
200 s binning of the light curves.  The black circles include the data from
the entire 20 ks observation.  The red squares are the F$_{var}$ recalculated,
neglecting the data between 10--15 ks (the X-ray flare).  The 6--12 keV data point in the 
second data set (red squares) is omitted due to degeneracy in the calculation
of F$_{var}$.
}
\label{fvar}
\end{figure}

As a first test for spectral variability we calculate the fractional variability amplitude
(F$_{var}$; Edelson et al. 2002) to examine the degree of variability in different energy bands.
The fractional variability amplitude is calculated in eight energy bins between 0.3--12 keV using
200 s binning of the light curves (Figure~\ref{fvar}).
The F$_{var}$ spectrum of I~Zw~1 is remarkably different from the typically flat spectrum
seen in other NLS1 such as 1H~0707--495 (Boller et al. 2002) and Ton~S180 (Vaughan et al. 2002).
The variations intensify as the energy increases and maximize between
2--12 keV.  The F$_{var}$ spectrum gives the sense of a variable hard power-law which
is physically distinct from the softer, less variable, emission component.  
 
Much of the spectral variability observed in Figure~\ref{fvar}
is associated with the X-ray flare seen in the broad-band light curve.  We recalculated the
F$_{var}$, but this time only for the low-flux state.  The result is
shown as the red squares in Figure~\ref{fvar}.  The low-flux state fractional variability in 
each bin is less than 10\%, and consistent with a constant across the
entire spectrum ($\chi^2$ = 2.6/6 $dof$).  
Figure~\ref{fvar} demonstrates that the spectral variability is stimulated by the 
flux outburst in Figure~\ref{lc}.

\subsection{The hard X-ray flare and flux-induced spectral variability}
\begin{figure}
\includegraphics[width=6.4cm,angle=-90,clip=]{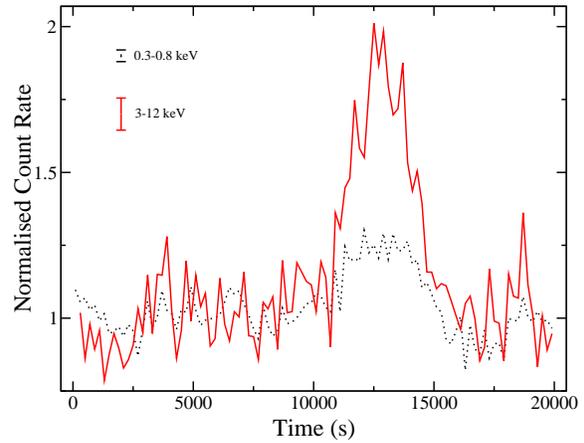}
\caption{The 0.3--0.8 keV (black dotted curve) and 3--12 keV (red solid curve) normalised light curves.
The vertical lines in the top left of the graph indicate the size of the average error bars.
Zero seconds on the time axis marks the start of the observation at 09:18:44 on 2002--06--22.
}
\label{normlc}
\end{figure}

By investigating the light curves in various energy bands we can demonstrate that the  
flare is concentrated at higher energies.  In Figure~\ref{normlc} we have plotted
the normalised light curves in the 0.3--0.8 keV and the 3--12 keV bands.  The light curves
are normalised to the average count rate during the periods 0--8.2 ks and 17--20 ks.
Clearly, the higher energy band is significantly more variable during the flaring event than the soft band.
During the flare the hard count rate increases
by $\sim$100\%, whereas the soft band shows only about a 20\% increase.  
This is the first time, to the best of our knowledge, that
a hard X-ray flare has been detected from a NLS1.

\begin{figure}
\includegraphics[width=6.4cm,angle=-90,clip=]{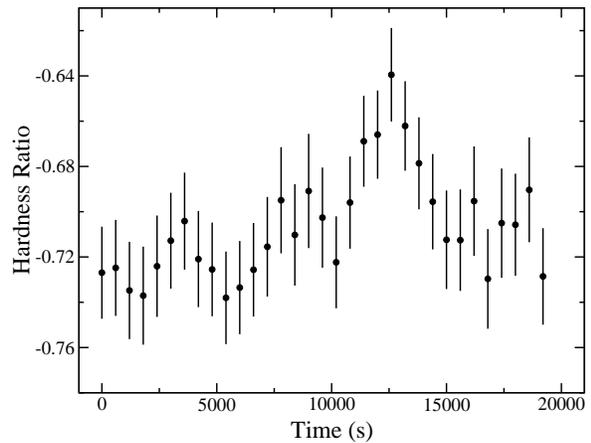}
\caption{Hardness ratio variability curve for $H$=2--12 keV and $S$=0.3--2 keV (see text for
details).  
Zero seconds on the time axis marks the start of the observation at 09:18:44 on 2002--06--22.}
\label{hr}
\end{figure}

The hardness ratio variability curve (Figure~\ref{hr}) is calculated using the formula        
$(H-S)/(H+S)$, where $H$ and $S$ are the count rates in the hard and soft bands, respectively.
The hardness ratio variability curve shows similar behavior as the light curve in
Figure~\ref{lc}.  The hardness ratio exhibits only small variations for most of the observation, 
but then
the fluctuations become enhanced between 10--15 ks, showing spectral hardening precisely during 
the time of the hard X-ray flare.
The significance of the relation between Figure~\ref{lc} and Figure~\ref{hr} is measured with the 
Spearman rank correlation coefficient to be significant at $>$ 5$\sigma$.

The fact that the flare is stronger at high energies implies that its origin is 
not in the cold accretion disc.  It is plausible that the flare originates in the accretion-disc
corona due to magnetic reconnection (e.g. Galeev et al. 1979).  The hard flare would then
irradiate the disc, where the high-energy photons will be reprocessed to lower energies via
Compton down-scattering.
In this simple picture, we may expect to see a time lag between the low and high energy 
light curves if the light travel time between the corona and disc is substantial.
In order to investigate this we calculate cross correlation functions using 200 s bins 
for a number of light curves in various energy ranges.
All of the cross correlations are relatively symmetric and consistent with zero time delay.
The results suggest that there are no leads or lags greater than $\pm$100 s between the various
energy bins.  Assuming the simple flare geometry discussed above, this result implies a very compact
system, in which the light travel time between the disc corona and the accretion disc itself
is $<$ 100 s.
All responses to the hard X-ray flare (light travel-time, reprocessing, spectral
changes) appear to occur on short time scales.

\begin{figure}
\includegraphics[width=6.4cm,angle=-90,clip=]{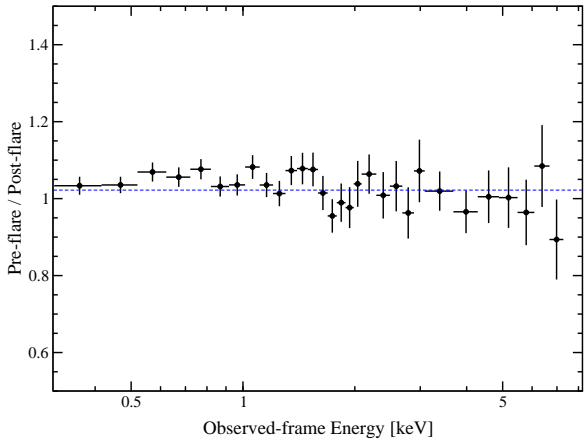} 
\caption{The ratio between the pre-flare and post-flare data.  The dotted line represents the 
ratio average.
}
\label{llratio}
\end{figure}

Another consideration is whether the flare, which appears to induce the spectral variability,
has any long-term 
effect on the spectrum; that is, even after the flux has returned to its pre-flare value.  To 
examine this possibility in a model-independent manner, we calculated the ratio between the pre-flare
data ($<$ 10 ks) and the post-flare data ($>$ 15 ks).  The ratio is displayed in Figure~\ref{llratio}.
There appears to be little difference in the ratio across the spectrum, and a constant fit to the ratio
is acceptable ($\chi^2$ = 29.5/30 $dof$).  Even the difference in the average flux is on the order 
of 2\%.  
However, on closer examination we notice that all nine data points below $\sim$1.2 keV lie above the
ratio average.  The probability that this is a random event is 1/512. Figure~\ref{llratio} indicates that
the soft flux prior to the flare is higher than the post-flare soft flux.  The shortage of post-flare data
makes it difficult to determine whether this subtle spectral difference is associated with the flare,
or a coincidental event.

\subsection{The nature of the spectral variability}

We have demonstrated that the flare is indeed hard, and that it induces
spectral variability.  The spectrum appears to become harder as the flare intensifies, and the changes 
brought on by the flare are immediate.
Could the spectral variability be
described by a component with a constant spectral slope but variable flux?  Or is it more likely that
the spectral variability is a result of a pivot in the X-ray spectrum at some higher energy?
Taylor et al. (2003) have presented a model-independent technique to distinguish between these forms
of spectral variability.  However, given the limited data we have available from this short observation
it is not possible to apply their technique successfully.

\begin{figure}
\includegraphics[width=6.4cm,angle=-90,clip=]{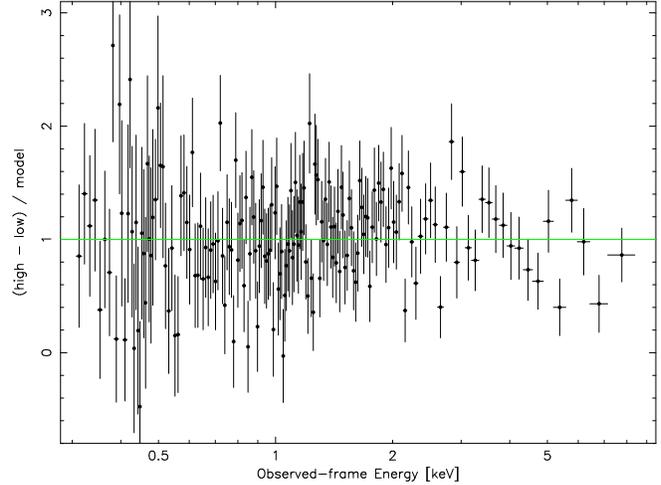}
\caption{The residuals from the difference spectrum (high-flux minus low-flux spectrum) 
fitted with a power-law modified by neutral absorption over the 0.3--10 keV band.  The data have been binned-up
for display purposes.
}
\label{diff}
\end{figure}

Fabian et al. (2002) and Fabian \& Vaughan (2003) have successfully used a difference spectrum (high-flux minus low-flux spectrum) 
to investigate the nature of the spectral variability in MCG--6--30--15.  
If the difference spectrum can be fitted with a power-law then the spectral variability between the two flux states 
can be 
attributed to a difference in
the normalisation of the intrinsic power-law without requiring a change in its spectral slope.
In Figure~\ref{diff}
we display the residuals from the difference spectrum for I~Zw~1 fitted with a power-law modified by cold absorption.  
The fit is a reasonable approximation ($\chi^{2}_{\nu}$ = 1.05). 

We further considered the spectral differences between the two flux states by modelling the high and low-flux state pn
spectra separately.  To both spectra we re-fit the double blackbody model described in Section~4.2 
(Table 1 model (a)).  
In both the low- and high-flux states, we found that the only significant
changes in the model parameters were to the normalisation of the continuum components. 
Within 90\% confidence the values for the temperatures of the thermal components, edge energy, and
neutral absorption were consistent.  
Even the photon indices of the low- and high-flux state power-law 
components were consistent ($\Gamma_{high}$ = 2.31$^{+0.06}_{-0.04}$ and $\Gamma_{low}$ = 2.35$^{+0.10}_{-0.06}$).
A noticeable difference occurred only in the relative fluxes ($f$ = F$_{3-10 keV}$ / F$_{0.3-2.0 keV}$) during
the two flux states.  During the low-flux state $f_{low}$ = 0.20 $\pm$ 0.01, whereas during the high-flux state
$f_{high}$ = 0.26 $\pm$ 0.02.

A further point to note is the possibility of changes in the strength of the iron line(s) between the 
different flux states.  The spectra during the two flux states show a very similar
continuum shape across the 0.3--10 keV band; however, the iron lines appear stronger during the low-flux (non-flare)
state while the normalisation of the lines remains consistent.  The equivalent widths of the neutral and ionised iron lines during the low state are
65$^{+4}_{-6}$ eV and 143$^{+10}_{-14}$ eV, respectively, while during the flare the strengths are
46$^{+2}_{-4}$ eV and 70$^{+4}_{-10}$ eV.  
Although the line parameters are especially model dependent and our exposure is short,
I~Zw~1 appears to be another object in which the iron lines do not vary as significantly or in response
to (in a simple manner) the continuum flux variations (Fabian et al. 2002; Lee et al. 2000, Reynolds 2000, 
Iwasawa et al. 1996).

\section{Summary}
We have presented a 20 ks {\em XMM-Newton} observation of I~Zw~1, the prototypical NLS1.
Our findings are summarized below.
\newline
\newline
(1)   The 0.3--10 keV continuum can be described by either a double blackbody plus a power-law or a broken
power-law, modified by intrinsic cold absorption.  The soft excess is weak, but notable
at energies between $\sim$0.4--0.7 keV.
Superimposed on this continuum are absorption and/or emission features below $\sim$ 1.0 keV,
and strong Fe K$\alpha$ emission between 6.4--6.8 keV. \\
(2)  The Fe K$\alpha$ emission can be successfully fitted with several models, and the current observation
does not allow us to determine the best one.  The pn data suggest that the 
Fe K$\alpha$ complex may be resolved into emission from a neutral line at 6.4 keV and an ionised component
at $\sim$6.9 keV (blend of He- and H-like Fe), weakening the case for a truly broad line.  \\
(3)  The low-energy spectrum can be modelled with either an absorption line or edge.
The current observation does not allow us to distinguish between these two models.  On the other hand,
a low-energy emission feature ($E$ $\approx$ 0.55 keV) can be introduced to the fit with relative success. \\
(4)  We find an X-ray flare which is isolated mostly in the 3--12 keV band.  We find that the
hard X-ray flare induces spectral variability in such a way that the spectrum becomes $harder$
during the flare.  There are no discernible lags between energy bands down to $\pm$100 s. 
Comparison of the pre-flare and post-flare spectra show that there is a subtle difference in the spectrum
prior to and after the flare.  It is not certain whether this spectral variability is related to the hard
flare.  \\
(5)  The flux of the iron line(s) do not appear to respond to the change in the continuum flux during
the flare.  The lines are strongest during the low-flux (non-flare) state.

\subsection{The low-energy spectra}
The quality of the data from this short exposure does not allow us to distinguish between
an absorption line or an edge, or an emission line in describing the features in the soft spectrum.
If the absorption is interpreted as a line then the best-fit energy may be suggestive of an UTA feature
due to Fe M absorption.  An absorption edge is a slightly better fit than the line.  The best-fit edge energy
is not consistent with that of OVII, which would indicate that the edge is severely blended with other
absorption features.  The emission line interpretation is consistent with emission coming from a blend
of OVII lines at $\sim$ 22 \AA.  The most likely scenario is that we have a combination of OVII emission
as well as absorption, but with the resolution of the EPIC instruments it is not possible to measure accurately 
the contribution from each component.
In any regard, we have demonstrated that the soft spectrum of I~Zw~1 is
complex and warrants a deep observation to utilise the RGS.

\subsection{The hard X-ray flare}
The detection of a hard X-ray flare, and spectral hardening during the flare, is entirely consistent
with theories of an accretion-disc corona which
is being heated by magnetic reconnection (e.g. Galeev et al. 1979).  The apparent instantaneous
response of the lower-energy emission mechanisms to the hard X-ray flare suggests that the
light travel-time between the emission regions is $<$ 100 s. 
Deeper analysis of the spectral variability suggests that spectral changes are due to an increase
in the high-energy flux relative to the low-energy flux, and not due to a change in the intrinsic shape of the continuum.
The weakness of the flare at low X-ray energies may be indicating that the seed photons for
the Comptonisation component are not coming from the 0.3--1 keV band, but probably from lower UV energies.
This observation could perhaps be useful in constraining Comptonisation models.
\newline

This short exposure with {\em XMM-Newton} has revealed a number of fascinating traits in I~Zw~1.
I~Zw~1 has once again proven to be an important object in allowing us to probe many interesting
facets of AGN, such as: X-ray flares, warm-absorption, and complex Fe K$\alpha$ emission,
A long exposure with {\em XMM-Newton} and {\em Chandra} would
be quite rewarding for studying these traits in greater detail. 

\begin{acknowledgements}
The authors are thankful to Frank Haberl for help regarding the EPIC calibration issues,
and the anonymous referee for helpful comments regarding the soft emission feature.
Based on observations obtained with {\em XMM-Newton}, an ESA science mission with
instruments and contributions directly funded by ESA Member States and
the USA (NASA).  WNB thanks NASA grants NAG5--9924 and NAG5--9933.
\end{acknowledgements}

\end{document}